\documentclass[aps,prl,twocolumn]{revtex4}
\usepackage{graphicx,epsf}
\usepackage{amsmath}
\usepackage{color}

\newcommand{\hsat}{h_{\rm sat}}

\begin{document}

\title{
Singular field response and singular screening of vacancies in antiferromagnets
}

\author{Alexander Wollny}
\author{Eric C. Andrade}
\author{Matthias Vojta}
\affiliation{Institut f\"ur Theoretische Physik, Technische Universit\"at Dresden,
01062 Dresden, Germany}

\date{\today}

\begin{abstract}
For isolated vacancies in ordered local-moment antiferromagnets we show that the
magnetic-field linear-response limit is generically singular: The magnetic moment
associated with a vacancy in zero field is different from that in a finite field $h$ in
the limit $h\to0^+$. The origin is a universal and singular screening cloud, which
moreover leads to perfect screening as $h\to0^+$ for magnets which display spin-flop
bulk states in the weak-field limit.
\end{abstract}

\maketitle


Defects are ubiquitous in solids. In magnets with localized spin moments, typical classes
of defects are missing or extra spins, arising, e.g., from substitutional disorder. Very
often, even small concentrations of such defects produce a large magnetic response at low
temperatures: Quasi-free spins cause a Curie tail in the magnetic susceptibility, which
then is routinely subtracted from raw experimental data. Assuming independent defects,
the amplitude of the Curie tail can be utilized to estimate the defect concentration,
provided that the behavior of a single defect is known.

Here we discuss the physics of isolated vacancies in antiferromagnets (AF) which display
semiclassical long-range order (LRO) in the ground state \cite{CHN}. In zero magnetic field, the state
with a single vacancy has a finite uniform magnetic moment, $m_0$, because the vacancy
breaks the balance between the sublattices. For collinear magnets, $m_0$ is quantized to
the bulk spin value, $m_0=S$ \cite{sandvik97,sbv99}, while in the non-collinear case fractional
values of $m_0$ occur due to the local relief of frustration \cite{wfv11}. These vacancy
moments are expected to show up in magnetization measurements, and they produce a
low-temperature Curie contribution to the uniform susceptibility in the two-dimensional
(2d) case where bulk order is prohibited by the Mermin-Wagner theorem
\cite{sbv99,sandvik03,sv03,sushkov03,wfv11}.

\begin{figure}[!b]
\includegraphics[width=0.4\textwidth]{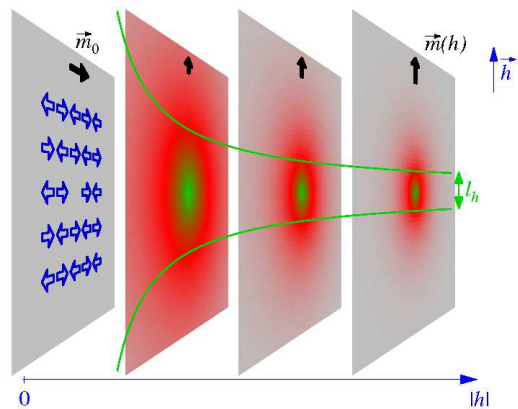}
\caption{
Illustration of singular screening for the square-lattice AF: At zero field, the vacancy
induces a moment $\vec{m}_0$ with $m_0=S$, which is modified into $\vec{m}(h)$ in the presence of a
uniform field $\vec{h}$. The size of the screening cloud, $l_h$, diverges as $h\to 0^+$,
and $m\to 0$ in this limit. For details see text.
}
\label{fig:ill}
\end{figure}

In this paper we show that, in an applied field $h$, non-trivial screening of the vacancy
moment occurs, such that the linear-response limit $h\to0^+$ is {\em singular} for a
magnet with a single vacancy, Fig.~\ref{fig:ill}:
The vacancy-induced magnetization jumps discontinuously from its zero-field value
$m_0$ to a different value $m(h\to0^+)$ upon applying an infinitesimal field $h$.
Thus, measurements of the vacancy-induced moment $m(h)$ in a finite field $h$ cannot
detect the zero-field value $m_0$ even for small $h$ \cite{symm_note},
which is of obvious relevance for any experiment trying to quantify
the defect contribution to a sample's magnetization or susceptibility.
Furthermore, the spin texture around the vacancy at finite $h$ has a piece \cite{eggert07} which is
singular as $h\to 0^+$ -- in a sense made precise below -- which screens the
vacancy-induced moment perpendicular to $\vec{h}$.
For magnets which feature spin-flop states (with all spins perpendicular to $\vec{h}$ as
$h\to 0^+$) in the absence of the vacancy, this leads to a semiclassical version of
{\em perfect screening} of the vacancy moment, $m(h\to0^+)=0$.

In the body of paper, we present general arguments and microscopic calculations
supporting these claims. Explicit results will be given in a $1/S$ expansion for
spin-$S$ AFs on 2d lattices, with
\begin{equation}
\label{H}
\mathcal{H} = \sum_{\langle ij\rangle}
J \vec{S}_i \cdot \vec{S}_j - h \sum_i S_i^z
\end{equation}
but our results are valid for AFs with LRO in any dimension $d$.
In particular, the singular response occurs for vacancies in the square-lattice AF, where
-- despite numerous studies
\cite{nagaosa,sandvik03,sushkov03,sushkov05,eggert06,eggert07,eggert11} -- it has been
overlooked to date \cite{prev_note}.
The singular behavior will be cut off for a finite vacancy density, and we
shall discuss the resulting crossover scales \cite{symm_note}.

\begin{figure}[!t]
\includegraphics[width=0.49\textwidth]{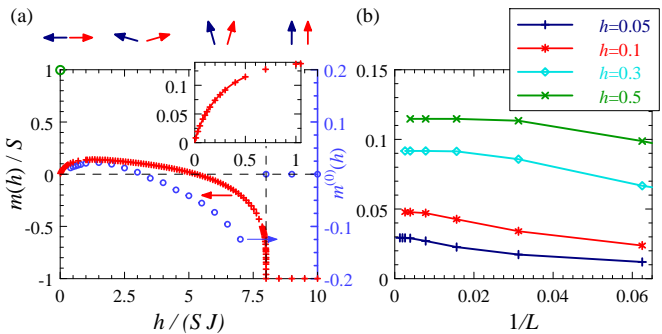}
\caption{
Vacancy contribution to the uniform magnetization, $m(h)$, for the square-lattice AF.
(a) Crosses: classical result $m(h)/S$, obtained on lattices with $L\leq768$ and extrapolated
to $L\to\infty$; the circle shows the linear-response value $|m_0/S|=1$.
Open dots: $1/S$ correction $m^{(0)}(h)$, here $L\leq 50$.
Vertical dashed line: saturation field $\hsat=8SJ$ \cite{sat_note}.
Top arrows: evolution of the bulk spin state with increasing $h$.
Inset: small-$h$ behavior of $m(h)/S$, together with a fit $m\propto h\ln
h$ (see text).
(b) Finite-size data $m(h)/S$ as function of $1/L$. $m$ saturates for $L>l_h$ with
$l_h\propto SJ/h$.
}
\label{fig:mh1}
\end{figure}

{\em Bulk behavior in a field.}
To set the stage, we recapitulate the evolution of a bulk AF state, which
spontaneously breaks the underlying SU(2) symmetry, upon application of a uniform field.
As $h\neq 0$ breaks the symmetry down to U(1), an infinitesimal field typically selects a subset
of states, which we refer to as $h\to 0^+$ bulk states:
For the square lattice, these are spin-flop states with staggered spin directions
perpendicular to $\vec{h}$. With increasing field the spins rotate toward the field
direction, until a fully polarized state is reached, see the illustration in
Fig.~\ref{fig:mh1}a.

\begin{figure}[!t]
\includegraphics[width=0.49\textwidth]{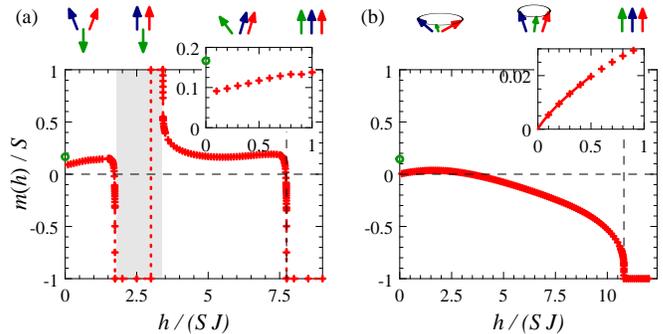}
\caption{
Vacancy-induced magnetization as in Fig.~\ref{fig:mh1}, but for the classical
triangular-lattice AF with biquadratic exchange of strength
(a) $k=-0.07$, leading to coplanar states in a field, and
(b) $k=0.1$, leading to umbrella states.
Here $\hsat=9SJ(1+2k)$, the shaded area in (a) indicates the magnetization plateau.
Insets: small-$h$ behavior.
Note that $m_0$ takes a $k$-dependent fractional value \cite{wfv11}.
}
\label{fig:mh2}
\end{figure}

For the triangular-lattice AF, an anomalously large ground-state degeneracy exists at the
classical level for $h\neq0$, which is lifted both by fluctuations and by additional
interactions \cite{kawa85,chub91,shannon11,laeuchli}.
Two cases are important: (a) ``coplanar'' and (b) ``umbrella'' states, see
Fig.~\ref{fig:mh2}. In (a) the spins are oriented in a plane in field direction, and the
evolution is via an intermediate up-up-down magnetization plateau (with a bulk
magnetization of 1/3 of the saturation value).
In contrast, in (b) the spins are arranged
in planar spin-flop states perpendicular to the field for $h\to 0^+$, and evolve in a
non-coplanar fashion continuously towards full polarization.
Below, we shall employ an additional biquadratic exchange $\sum_{\langle ij\rangle} K (\vec{S}_i \cdot
\vec{S}_j)^2$ with relative strength $k=K S^2/J$ to select between the two cases in a
classical calculation: $k>0$ ($k<0$) favors umbrella (coplanar) states, with $|k|<2/9$
required to preserve the familiar 120$^\circ$ order at zero field \cite{laeuchli}.


{\em Numerical results.}
We now present numerical results for the single-vacancy finite-field ground state of
square and triangular AFs. We consider the vacancy contribution to the magnetization,
$m(h)$, defined as the difference between the total magnetizations of the system with and
without vacancy.
Figs.~\ref{fig:mh1} and \ref{fig:mh2} display $m(h)$ for the classical AF on the square
and triangular lattices and demonstrate our main results:
(i) $m(h\to 0^+)$ does not reach the zero-field moment $|m_0|$ in any of the cases,
i.e., the impurity magnetization jumps upon application of an infinitesimal field.
(ii) For the square lattice we find $m(h\to 0^+)\to 0$; the same happens for umbrella
states in the triangular lattice. In contrast, $m(h\to 0^+)$ tends to a finite value in the
coplanar triangular-AF case.
Fig.~\ref{fig:mh1}a also shows the next-to-leading term in a $1/S$ expansion for $m(h)$,
indicating that quantum corrections do not qualitatively change these results.

In the remainder of the paper we explain the physics behind these striking observations.
As a first step, we show that the spin configurations are necessarily different for $h=0$
and $h\to0^+$ in the presence of a vacancy.


{\em Vacancy: Zero-field state.}
The vacancy breaks the balance between the sublattices of the host AF and locally
distorts the bulk state.
If the AF is collinear (and remains collinear upon introducing the vacancy) there is no
distortion in the classical limit, and the vacancy-induced magnetization is $m_0=S$.
Quantum fluctuations arise from the action of $S_i^+ S_j^-$ terms in the Hamiltonian and
hence conserve total spin, such that $m_0$ remains locked to $S$; however, the amplitudes
of the $\langle \vec{S}_i\rangle$ are modified near the vacancy.
In non-collinear AFs, the directions of the $\langle \vec{S}_i\rangle$ re-adjust in
response to the vacancy because frustration is locally relieved. Then,
$m_0$ takes a fractional value which depends on $S$ and microscopic details \cite{wfv11},
i.e., the vacancy spin gets partially screened, and both
undercompensation and overcompensation are possible.

In all cases, the local distortions decay algebraically with the distance $r$ to the
vacancy due to the presence of Goldstone modes. Typically this decay follows $1/r^{d+1}$
\cite{sushkov05,wfv11}, such that the total (i.e. integrated) vacancy contribution to any
observable is bounded, i.e., of order unity -- this we refer to as a ``regular''
screening cloud.


{\em Singular $h\to 0^+$ limit.}
Now consider applying an infinitesimal field $\vec{h}$. To this end, rotate the $h=0$
state in spin space such that the spins far away from the vacancy match a $h\to0^+$ bulk
configuration. This generates a state with finite magnetization $\vec{m_0}$
which, however, is in general {\em not} oriented in field direction. This is illustrated
in Fig.~\ref{fig:ill} for the square lattice, where the spins are oriented perpendicular
to the infinitesimal field -- the same applies to $\vec{m_0}$. Clearly, such
a state {\em cannot} be the ground state for $h\neq 0$, as stability demands $\vec{m}
\parallel \vec{h}$ in the presence of SU(2) symmetry.
In other words, the states at $h=0$ and $h\to0^+$ are required to differ because the
latter is subject to the additional constraint that the uniform moment has to point in
field direction (while both states have matching spin directions far away from the
vacancy). Fulfilling this constraint requires an additional distortion to screen the
moment perpendicular to $\vec{h}$.

This analysis also reveals exceptions {\em without} zero-field singularity in the
presence of a vacancy: This happens when the properly rotated zero-field state has a
moment $\vec{m_0}$ which points along the direction of the infinitesimal $\vec{h}$. A
concrete example is the triangular lattice with coplanar states and
undercompensated impurity moments \cite{wfv11}. (In the classical limit, there is
overcompensation for all $k<0$ \cite{wfv11} which leads to results as in
Fig.~\ref{fig:mh2}a, but undercompensation can be expected for $S<\infty$.)

We note that mean-field theory, neglecting spatial variations and thus distributing the
screening process over the entire system, qualitatively reproduces the zero-field
singularity in $m(h)$, but fails in other respects \cite{suppl}.


{\em Finite-field screening cloud.}
We turn to the finite-field screening, in order to understand the difference
between the cases in Figs.~\ref{fig:mh1}a and \ref{fig:mh2}a, where $m(h\to 0^+)$ reaches
zero or a finite value, respectively.
The vacancy-induced modification of the directions of $\langle\vec{S}_i\rangle$ can be
parameterized by the spherical angles $\delta\Theta_i$, $\delta\phi_i$ (one angle is
sufficient for the coplanar cases in Figs.~\ref{fig:mh1}a and \ref{fig:mh2}a).
Considering that the distortion serves to screen the vacancy moment's component perpendicular to
$\vec{h}$, the screening can be quantified using
\begin{equation}
\vec{m}_\perp(r) = \sum_{|\vec{r}_i|\leq r} \langle\vec{S}_i\rangle_{\perp}
\label{mperp}
\end{equation}
where $\langle\vec{S}_i\rangle_{\perp}$ is the moment perpendicular to $\vec{h}$, and the
sum runs over all sites with a distance to the vacancy closer or equal $r$. Clearly,
$\vec{m}_\perp(r\to\infty)$ is the total magnetization perpendicular to $\vec{h}$ and has
to vanish.

\begin{figure}[!t]
\includegraphics[width=0.49\textwidth]{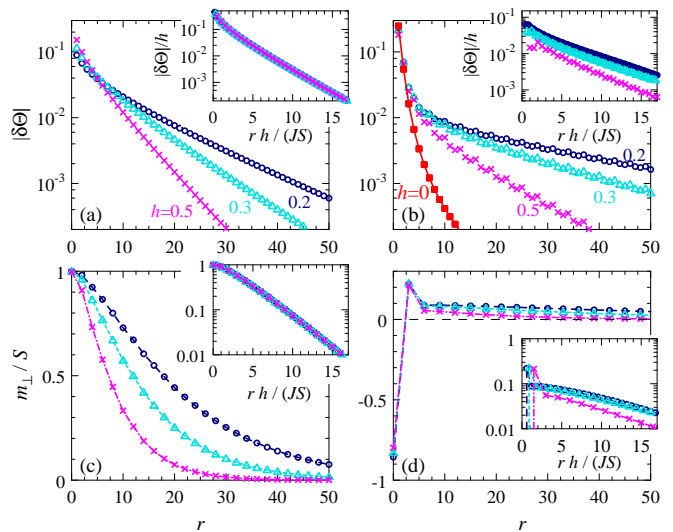}
\caption{
Field-induced screening cloud for (a,c) the square lattice and (b,d) the triangular
lattice with $k=-0.07$, for different values of $h$.
(a,b) Vacancy-induced rotation angles $\delta\Theta(\vec{r})$ as function of the distance
$r$ to the vacancy. In (b) the zero-field angles are shown as well. Insets:
scaling of $\delta\Theta/h$ vs. $rh/(SJ)\propto r/l_h$, with the short-range
piece projected out in (b).
(c,d) Transverse magnetization $\vec{m}_\perp(r)$ as function of the integration radius
$r$ in Eq.~\eqref{mperp}. Insets: $\vec{m}_\perp$ vs. rescaled distance $rh/(SJ)$.
}
\label{fig:scr1}
\end{figure}

We start with the square lattice: The angles $\delta\Theta$, Fig.~\ref{fig:scr1}a,
decay exponentially on a length scale $l_h \propto SJ/h$. This is expected: The
distortion is mediated by a bulk mode which is a gapless Goldstone mode in zero field,
but acquires a gap $\propto h$ for $h>0$. A remarkable feature is that
$\delta\Theta_i$ at fixed $\vec{r}_i$ has a non-monotonic field dependence and appears to
reach zero as $h\to 0$ (visible at small $r$). This implies that the
distortion, and with it the screening, happens at progressively larger $r$ with
decreasing $h$.
This is confirmed by the plot of $\vec{m}_\perp(r)$, Fig.~\ref{fig:scr1}c, showing that
the cloud size increases as $h\to 0^+$.

To understand these results analytically, we resort to a continuum description: The
small-field distortions consist of smooth variations of the order-parameter field
$\vec{\varphi}(\vec{r})$ (or, equivalently, of the angles $\Theta$ on all sublattices). For
a single vacancy, those distortions have been analyzed in Ref.~\onlinecite{eggert07}, and
found to decay as $h^{d-1}/(\rho_s c^{d-2}) f_d(hr/c)$ where $\rho_s$ and $c\propto SJ$ are the
spin stiffness and spin-wave velocity, respectively, and $f_d$ is a universal
function which depends on the dimension $d$ only. Specifically, $f_2(x) = K_0(x)$ (the
modified Bessel function), and $f_3(x)=\exp(-x)/x$.

To analyze the screening of $m_\perp$, we first note that distortion-induced (local)
contributions to $m_\perp$ arise -- in the presence of a finite bulk magnetization
density $\vec{l} \propto \vec{h}$ -- to linear order in $\delta\Theta$, $\delta m_\perp
\propto \int d^d r |\vec{l}| \delta\Theta$, which have to compensate the bare
vacancy-induced transverse moment (which itself equals $S$ in the square-lattice case).
Using the above long-distance form of the distortion, we have
$\delta m_\perp \propto (c^2/\rho_s) (h/c)^d \int d^dr f_d(hr/c) \propto
(c^2/\rho_s) \int d^dx f_d(x)$. As the last integral is finite, we see that all $h$
dependence has dropped out, indicating stable screening of $m_\perp$ for any
$h$.

The insets in Figs.~\ref{fig:scr1}a,c confirm the anticipated scaling of our data with
$l_h$: Plotting $\delta\Theta/h$ and $\vec{m}_\perp$ each as function of rescaled
distance $rh/(SJ)$ reveals data collapse, i.e., the screening cloud is {\em universal}
for small fields. The scaling of $\delta\Theta/h$ also implies that the texture does
not approach a well-defined limiting form as $h\to 0^+$, but instead has a diverging size
combined with a vanishing amplitude -- this we refer to as ``singular'' screening cloud.

A similar calculation can be employed to obtain the cloud's contribution to the
field-parallel magnetization $m_\parallel\equiv m(h)$. This arises from two higher-order
terms, $\int d^dr |\vec{l}| \delta\Theta^2$ and $\int d^dr |\vec{\varphi}|\vec{\nabla}^2\Theta$, while
lower orders vanish because the transverse magnetization density is zero in the
bulk reference state. The first term scales as $h^{d-1} \int d^dx f_d^2(x)$ while the
second one scales as $h \int d^dx f_d''(x)$. Importantly, the first integral is regular
whereas the second one has a short-distance logarithmic divergence in any $d$ -- this is cut off
by noting that the lower limit of the $x$ integral is not zero, but $\propto h$.
As a result, the cloud's contribution to $m(h)$ follows $(h \ln h)$ for small $h$.
Indeed, our numerical results in Fig.~\ref{fig:mh1}a are perfectly fit by this
form -- note that the bare vacancy-induced moment along $\vec{h}$ vanishes $\propto h$
and is subleading.

Now we turn to the triangular lattice. An analysis of the angles
$\delta\Theta(r)$, Fig.~\ref{fig:scr1}b, shows that they arise from a superposition of two
textures, with spatial $s$-wave and $f$-wave symmetries, respectively. The latter is well
localized near the vacancy and evolves smoothly into the zero-field texture
\cite{wfv11}, while the former is a field-induced distortion very similar
to that in the square lattice.
As a result, we find two-stage screening at small $h$: At short distances, the moment of
the missing spin is screened to $\vec{m}_0$ in a regular and weakly field-dependent fashion, while
the field-perpendicular component of this $\vec{m}_0$ is then compensated at larger
distances via a singular cloud of size $l_h$.
This singular piece can be isolated by projecting $\delta\Theta(r)$ onto its $s$-wave component,
Fig.~\ref{fig:scr1}b, and it admits the same continuum description as above. Because of the
nearby magnetization plateau, universal scaling is reached at much smaller $h$ as
compared to the square lattice.

Finally, this insight allows to deduce the value $m(h\to 0^+)$:  This is simply the
field-parallel part of $\vec{m}_0$ in the (properly rotated) zero-field state, because
the contribution of the singular screening cloud vanishes as $(h\ln h)$. For the
triangular lattice with overcompensated impurity, this $\vec{m}_0$ forms a $60^\circ$ angle
with $\vec{h}$ \cite{wfv11} such that $m(h\to 0^+) = |\vec{m}_0|/2$, matching our numerical
result in Fig.~\ref{fig:mh2}a, while for all systems with spin-flop states $m(h\to
0^+)=0$, i.e., the vacancy moment is perfectly screened.


{\em Quantum effects.}
Although the concrete calculations so far were for classical spins, the results
qualitatively apply to any system with LRO: As happens in general for non-collinear
magnets, quantum effects will modify directions and amplitudes of $\langle
\vec{S}_i\rangle$. However, the existence of both the zero-field singularity and the
perfect screening are unaffected, as they derive from symmetry and stability arguments.

To illustrate this, we have calculated quantum corrections to $m(h)$ in a $1/S$ expansion
using spin-wave theory on finite lattices \cite{wessel,nikuni,error_wfv11}.
A sample result for the square-lattice
$S^0$ term of $m(h)$ is shown in Fig.~\ref{fig:mh1} \cite{sat_note} -- its $h\to 0$ behavior is
consistent with $(h\ln h)$.


{\em Limits and crossover scales.}
The zero-field singularity only occurs in the thermodynamic limit \cite{symm_note}:
$m$ evolves smoothly for a single vacancy in a finite-size
system, and the limits $N\to\infty$ and $h\to 0$ do not commute \cite{suppl}.

Assuming now a finite vacancy concentration $n_{\rm imp}$, the singularity is
replaced by a crossover governed by the two length scales $l_h$ and $l_{\rm imp} = n_{\rm
imp}^{-1/d}$, the mean vacancy distance. For $l_h\ll l_{\rm imp}$, the vacancy moments
respond independently to the field, whereas for $l_h\gg l_{\rm imp}$ the screening clouds
overlap and hence their moments tend to average out (for an equal distribution over
all sublattices). Thus, the magnetization per vacancy will follow $m(h)$ as calculated
here for elevated fields, but will (smoothly) vanish below a crossover field given by
$h/J \sim n_{\rm imp}^{1/d}$.

These considerations also apply at finite temperatures inside the ordered phase of 3d
Heisenberg magnets. In 2d, order occurs only at $T=0$, but the bulk correlation length
$\xi$ becomes exponentially large as $T\to 0$. Then, the physics is governed by the
interplay of the three length scales $l_h$, $l_{\rm imp}$, and $\xi$, and we quickly
discuss some interesting limits. For $\xi\gg l_h,l_{\rm imp}$ one recovers the $T=0$
physics discussed above. Single-impurity physics obtains in the dilute limit, $l_{\rm
imp}\gg \xi,l_h$, where at elevated fields, $\xi\gg l_h$, the magnetization per vacancy
again follows our $m(h)$. In contrast, in the low-field limit, $\xi\ll l_h$, linear response
is restored such that the vacancy susceptibility takes the Curie form $\chi=m_0^2/(3kT)$ \cite{sbv99,sandvik03,sv03,sushkov03,wfv11},
i.e., the singularity is replaced by a crossover on the scale $h/J\sim\exp(-T/J)$.
Finally, the limit $l_h\gg\xi,l_{\rm imp}$ is governed by zero-field physics.


{\it Conclusions.}
For a single vacancy in an ordered AF we have found that the magnetic behavior is
generically singular in the weak-field limit: The vacancy contribution to the
magnetization jumps discontinuously upon applying an infinitesimal field. The singularity
can be traced back to non-trivial field-induced screening due to a universal and singular
screening cloud in the $h\to 0^+$ limit.

Our predictions can be experimentally verified in any AF doped with a small, controlled
amount of vacancies, provided that magnetic anisotropies are small \cite{symm_note}. In
addition, numerical studies beyond the $1/S$ expansion (e.g. quantum Monte Carlo) are
called for.


We thank J. Kunes, R. Moessner, and O. I. Motrunich for discussions, and L. Fritz for a
collaboration at an early stage of this work. This research was supported by the DFG
through FOR 960 and GRK 1621.



\begin{thebibliography}{}
\bibitem{symm_note}
Violations of SU(2) symmetry remove the zero-field singularity, but only finite-$h$ measurements
along an easy axis can possibly the zero-field vacancy moment $m_0$.
\bibitem{prev_note}
The $h\to0$ singularity was observed numerically in Ref.~\onlinecite{wfv11} for a
specific triangular-lattice case, but the generality and the underlying physics were not
understood there.
\bibitem{sat_note}
As $h\to\hsat^-$, the classical-limit $m(h)$ is continuous with an infinite slope at
$\hsat$, associated with a diverging correlation length. Fluctuation corrections, already
logarithmically singular without vacancy \cite{nikuni}, require a more detailed analysis which is
beyond the scope of this work.
\bibitem{error_wfv11}
For the triangular lattice at $k=0$ we find $m_0=-0.04S + 0.35 + \mathcal{O}(1/S)$, which
corrects our earlier result in Ref.~\onlinecite{wfv11}, where we missed the $1/S$
correction to the angles $\Theta_i$ \cite{nikuni}.
\end{thebibliography}
\end{document}


\title{
Supplemental material for:\\
``Singular field response and singular screening of vacancies in antiferromagnets''
}

\author{Alexander Wollny}
\author{Eric C. Andrade}
\author{Matthias Vojta}
\affiliation{Institut f\"ur Theoretische Physik, Technische Universit\"at Dresden,
01062 Dresden, Germany}

\date{\today}

\maketitle


\section{Vacancy in a finite-size system}

While the analysis in the main text was concerned with the vacancy magnetism in an
infinitely large host, we discuss here the behavior for a finite host of size $N=L^d$
where $L$ is the linear dimension, i.e., where the vacancy-doped system has $(N-1)$ spins.

The zero-field singularity advertised in the paper is due to competing boundary (i.e.
impurity) and bulk magnetization processes, described by terms in the system's energy of
the form $-h m$ and $-\chi h^2 N$, respectively, where $\chi$ is the bulk susceptibility.
Importantly, the limits $N\to\infty$ and $h\to 0$ do not commute:
If $h$ is taken to zero in a finite-size system, then the impurity contribution to the
field energy will always dominate over the bulk contribution for sufficiently small $h$,
such that the zero-field vacancy moment $m_0$ simply becomes field-aligned for $h\ll
SJ/N$. In this limit, the bulk spins will not have taken their ``spin-flop''
configuration. Upon increasing $h$, the bulk term gradually starts to compete with the
impurity term, leading to a smooth evolution of $m$ in a finite-size system, with a crossover
between small-field and large-field behavior at $h\sim SJ/N$.

In contrast, in the thermodynamic limit where $N\to\infty$ is taken first, the bulk
contribution to the field energy dominates for any finite $h$, such that the bulk spins
are in their ``spin-flop'' configuration which forces $m(h)$ to be different from $m_0$
as discussed in the main text. In other words, the above finite-size crossover in $m(h)$
becomes the advertised jump occurring at $h=0^+$ in the thermodynamic limit.

\begin{figure*}[!t]
\includegraphics[width=0.61\textwidth]{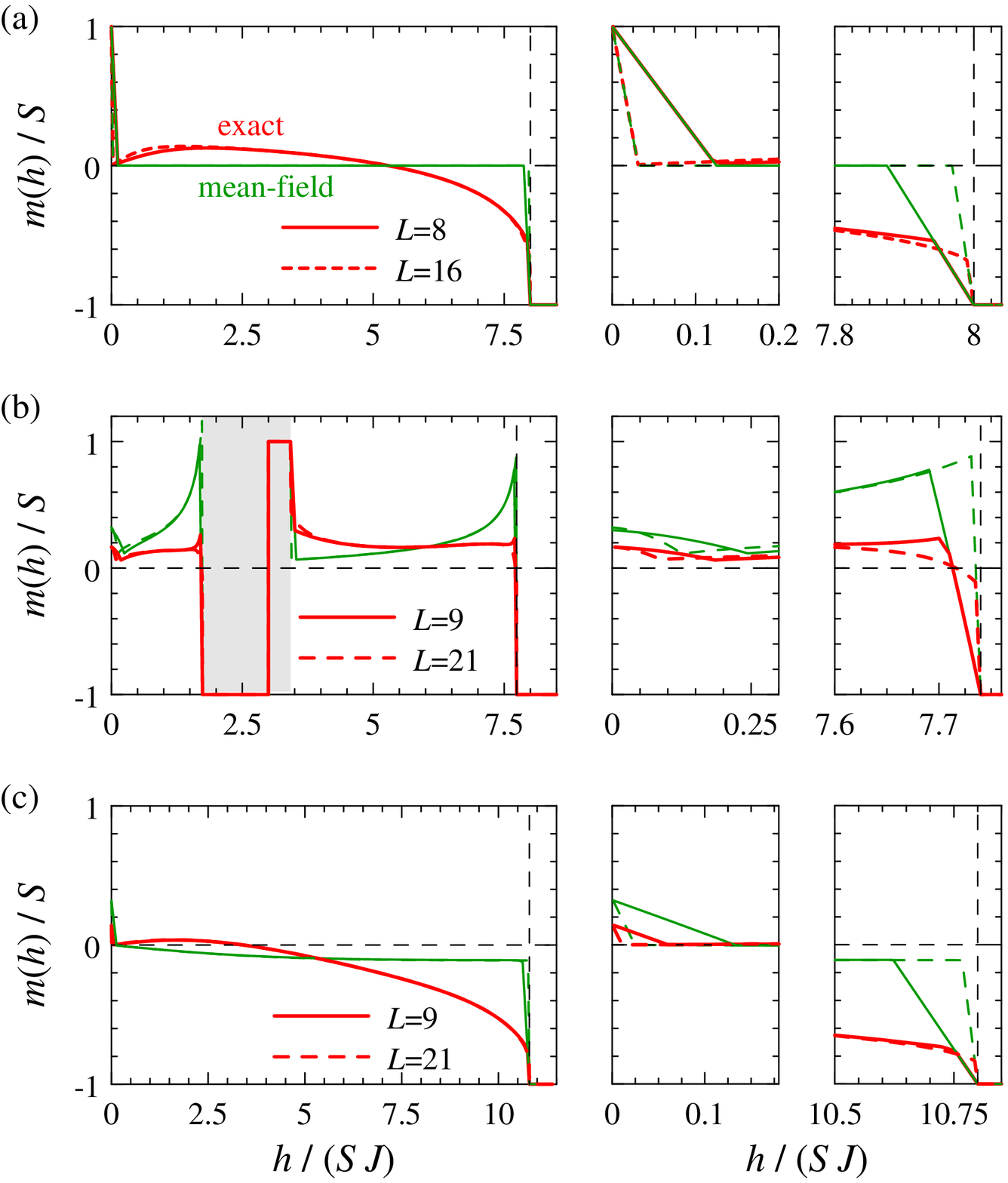}
\caption{
Vacancy contribution to the uniform magnetization, $m(h)$ (red), calculated for a classical
finite system of size $N=L^2$, for
(a) the square-lattice AF,
(b) the triangular-lattice AF with $k=-0.07$, and
(c) the triangular-lattice AF with $k=0.1$.
In all cases, the left panels show the full field range, while the smaller right panels
magnify the behavior near $h=0$ and $h=\hsat$.
The corresponding infinite-system results are in Figs. 2 and 3 of the main paper;
the shaded region in panel (b) indicates the magnetization plateau.
Also shown are the results for $m(h)$ from the mean-field approach (green) as given by
Eqs.~(\ref{mf1},\ref{mf2}) for the same parameters and system sizes, for details see
text.
}
\label{fig:S1}
\end{figure*}

These considerations are well borne out by numerical simulations on small systems in the
classical limit, Fig.~\ref{fig:S1}, where $m(h)$ follows an initial decrease from $m_0$
for small $h$ before it sharply crosses over into its $N\to\infty$ behavior for larger
$h$. For the square lattice, the crossover field is given by $h_x = 8SJ/N$.

We note that this finite-size crossover is irrelevant for experiments in macroscopic
solids, because there any applied field is larger than $J/N$. Instead, one needs to
consider the crossovers induced by a finite concentration of vacancies, which are
discussed toward the end of the main text. The finite-size crossover may, however, be
probed in finite systems realized, e.g., with ultracold gases in optical lattices.

\section{Mean-field approximation for a vacancy-doped magnet}

For completeness, we describe here the results of a mean-field approach to the
(classical) vacancy-doped antiferromagnet. The simplest approximation assumes all spins on
a given sublattice to be parallel, such that each sublattice can be represented by a
single ``superspin''. For the square lattice, this amounts to a mean-field energy of the
form
\begin{equation}
E_{\rm MF} = \frac{2zJ}{N} \vec{S}_A \cdot \vec{S}_B - h (S_A^z + S_B^z)
\label{mf1}
\end{equation}
where $z=4$ is the number of nearest neighbors and $\vec{S}_{A,B}$ represent the
magnetization vectors of the sublattices $A$ and $B$, with $|\vec{S}_A| = SN/2$ and $|\vec{S}_B| =
S(N/2-1)$ for the single-vacancy case.
Minimizing $E_{\rm MF}$ w.r.t. the angles parameterizing the directions of $\vec{S}_{A,B}$
yields a mean-field approximation, e.g., for the field-dependent magnetization of the
vacancy-doped antiferromagnet from which one can obtain the vacancy-induced magnetization
$m(h)$ as defined before.

Formally, Eq.~\eqref{mf1} is a mean-field theory for a ferrimagnet;\cite{clark} any spatial
variations which are expected due to the presence of the impurity are discarded. Such a
mean-field theory has been briefly considered for the vacancy problem in
Ref.~\onlinecite{eggert07}.

Mean-field results for $m(h)$ for finite square lattices are included in Fig.~\ref{fig:S1}a.
As compared to the numerically exact results, various features are apparent:
\begin{itemize}

\item The mean-field results are extremely accurate for small fields $h<8SJ/N$. This can be
rationalized by considering that, in the full solution, the field-induced length scale of
the spin texture, given by $l_h \propto SJ/h$, then becomes $l_h>N$ in this
regime. Hence, $l_h\gg L$, which implies that there is essentially no spatial variation of
the spin pattern across the entire system, and consequently a theory neglecting spatial
variations is accurate.

\item The mean-field theory incorrectly predicts $m(h)=0$ for all intermediate fields $8/N
<h/(SJ) < 8-8/N$.

\item Taking the limit $N\to\infty$, the mean-field theory reproduces the correct jump in $m(h)$
at zero field, but incorrectly predicts a jump at $h=\hsat^-$ as well.

\end{itemize}

For the triangular lattice, the mean-field approach can be generalized to three sublattices, where
now
\begin{equation}
E_{\rm MF} = \frac{3zJ}{2N} (\vec{S}_A \cdot \vec{S}_B + \vec{S}_A \cdot \vec{S}_C + \vec{S}_B \cdot \vec{S}_C) - h (S_A^z + S_B^z + S_C^z)
\label{mf2}
\end{equation}
with $z=6$ and $|\vec{S}_{A,B}| = SN/3$ and $|\vec{S}_C| = S(N/3-1)$ for a single
vacancy. An additional biquadratic interaction (see main text) is straightforward to include.

Figs.~\ref{fig:S1}b,c show the mean-field results for $m(h)$ in the triangular lattice
with $k=-0.07$ and $k=0.1$ as in the main paper, vis-a-vis the exact results.
While the mean-field theory again predicts a jump of $m(h)$ at $h=0^+$ in the $N\to\infty$
limit, its deficiencies are even more apparent than in the square-lattice case:
The mean-field theory fails to properly capture the non-singular piece of the screening process,
such that both $m_0$ and the jump height are incorrect. Further, large deviations occur
near the magnetization plateau and the saturation field.

In summary, the mean-field approach qualitatively reproduces the small-field finite-size
crossover in $m(h)$ and its zero-field jump in the thermodynamic limit, essentially
because it is bound to follow the general arguments laid out in the main text. However,
the neglect of spatial fluctuations renders the mean-field theory unreliable for various
important aspects of the impurity problem at hand.
